\documentclass[twocolumn,amsmath,amssymb,prl]{revtex4}
\usepackage{latexsym}
\usepackage{graphicx}
\usepackage{psfig}
\usepackage{color}
\usepackage{verbatim}

\usepackage[letterpaper, margin=1in]{geometry}
\begin{document}

\title{Andreev reflection of quantum Hall states through a quantum point contact}
\author{Mehdi~Hatefipour$^{1}$}
\author{Joseph J. Cuozzo$^{2}$}
\author{Ido~Levy$^{1}$}
\author{William M. Strickland$^{1}$}
\author{Dylan~Langone$^{1}$}
\author{Enrico Rossi$^{3}$}
\author{Javad~Shabani$^{1}$}
\email{jshabani@nyu.edu}

\affiliation{
$^{1}$Center for Quantum Information Physics, Department of Physics, New York University, NY 10003, USA\\
$^{2}$Materials Physics Department, Sandia National Laboratories, Livermore, CA 94551, USA\\
$^{3}$ Department of Physics, William \& Mary, Williamsburg, Virginia
23187, USA 
}
\date{\today}

\begin{abstract}
 We investigate the interplay between the quantum Hall (QH) effect and superconductivity in InAs surface quantum well (SQW)/NbTiN heterostructures using a quantum point contact (QPC). We use QPC to control the proximity of the edge states to the superconductor. By measuring the upstream and downstream resistances of the device, we investigate the efficiency of Andreev conversion at the InAs/NbTiN interface. Our experimental data is analyzed using the Landauer-Büttiker formalism, generalized to allow for Andreev reflection processes. 
 We show that by varying the voltage of the QPC, $V_{QPC}$, the average Andreev reflection, $A$, 
 at the QH-SC interface can be tuned from 50\%
 to $\sim 10$\%. 
The evolution of $A$ with $V_{QPC}$ extracted from the measurements exhibits plateaus separated
by regions for which $A$ varies continuously with $V_{QPC}$.
The presence of plateaus suggests that for some ranges of $V_{QPC}$ 
the QPC might be pinching off almost completely 
 from the QH-SC interface some of the edge modes.
Our work shows a new experimental setup to control and advance the understanding 
of the complex interplay between superconductivity and QH effect in two-dimensional electron gas systems.
%
\end{abstract}

\maketitle

\section{\textbf{I. Introduction}}
In recent years, the realization of topological phases of matter has been a focus of intense research in the field of condensed matter physics~\cite{Kitaev_2001,Beenakker2013,AliceaReview}. This is partly motivated by the potential of such systems to host exotic quasiparticles with non-Abelian statistics, which could be used for fault-tolerant quantum computation through braiding operations~\cite{Nayak2008,Laucht2016,AliceaReview}. While InAs surface quantum wells (SQWs) in proximity to superconductors have emerged as a promising platform for the realization of topological superconductivity~\cite{microsoft2022,Matthieu2021,Lee2017}, it is proposed that combining rich quantum Hall physics with superconductivity can allow access to higher topological states~\cite{Meng14,Clarke13}.

Recent studies~\cite{Mehdi2022,Gul2022, Lee2017, Zhao2020, Zhao2022, Amet2016} have demonstrated the potential of proximitizing quantum Hall edge states and superconductivity. While these studies are performed on various platforms the signature of Andreev reflection is observed through negative downstream resistance and a reduction in Hall (upstream) resistance in the quantum Hall regime. Specifically in Ref. ~\cite{Mehdi2022}
we showed that InAs/NbTiN systems can exhibit up to 60\% Andreev conversion. This is attributed to the high efficiency of Andreev conversion at the InAs/NbTiN interface, resulting from the strong hybridization of the quantum Hall edge modes with the states in the superconductor. While cleanliness of the interface is crucial in achieving 
high Andreev conversion, the underlying microscopic understanding of this negative resistance is still not well-understood. These recent experimental works have galvanized theoretical efforts to study Andreev processes in quantum Hall edge state transport involving superconductivity~\cite{Tang2022, David2022, Kurilovich2023, Michelsen2023, Kurilovich2022_criticality, Manesco2022, Beconcini2018, Alavirad2018, Schiller2022, Galambos2022, Cuozzo2023}.
In addition to the studies mentioned above, recent study \cite{vignaud2023evidence} has shown evidence for chiral supercurrent in quantum Hall Josephson junctions, and Ref. ~\cite{uday2023induced} has reported the observation of crossed Andreev reflection (CAR) across a narrow superconducting Nb electrode contacting the chiral edge state of a quantum anomalous Hall insulator (QAHI). These studies provide concrete demonstrations of the hybridization of superconductivity and quantum Hall effects.

In this work, we introduce a new tool that is often used in studying mesoscopic features of quantum Hall physics \cite{Moty2009,Gabelli2007,Sanchez2020,Taktak2022,Baer2014,Zimmermann2017,Nayak2008} to control the number of edge modes that interact with the superconductor.
To achieve this we place a voltage-controlled constriction, quantum point contact (QPC) \textit{exactly} on the interface between InAs and NbTiN to control the interplay between the QH edge modes and the superconductor.

\section{\textbf{II. sample growth and preparation}}

\begin{figure*}[t]
\centering
\includegraphics[width=1\textwidth]{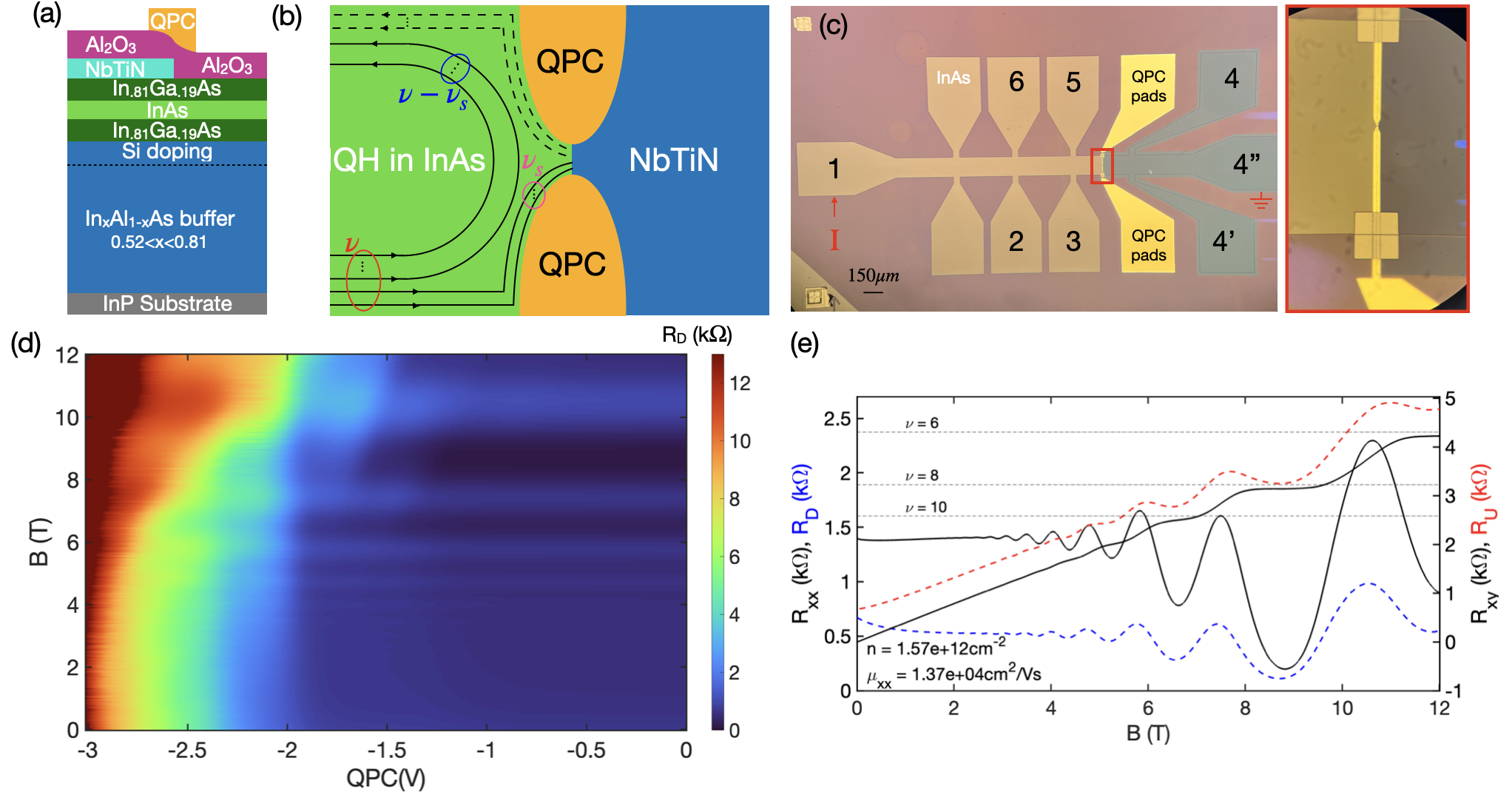}
\caption{(a) Schematic of the grown and fabricated layers of the sample (the region of the device specified by the red rectangle in panel (c)). (b) Schematic diagram of the device showing the InAs/NbTiN interface with a QPC in the integer quantum Hall regime. The solid and dashed lines represent electrons and holes, respectively. When the QPC is activated, edge states allowed to pass through the QPC can be Andreev reflected off the NbTiN interface. (c) An optical photo of the finished device, highlighting the different materials labeled by color and the contacts used for current source and drain. (d) Downstream resistance (R$_D$) as a function of perpendicular magnetic field and QPC voltages. R$_D$ is measured between contacts 5 and 4. (e) The 2DEG magnetotransport data of the fabricated device, where R$_{xx}$ and R$_{xy}$ are measured between contacts 6-5 and 5-3, respectively. The plateaus are labeled by their corresponding filling factors. Additionally, we show R$_D$ (between contacts 5 and 4) and R$_U$ (between 3 and 4') plots with blue and red color, respectively.}
\label{fig:fig_1_cartoon}
\end{figure*}

The semiconductor in our study is a molecular beam epitaxy (MBE) 
heterostructure grown on a semi-insulating InP(100) wafer. To form a two-dimensional electron gas (2DEG), a quantum well (QW) was grown on an In$_{x}$Al$_{1-x}$As buffer with a graded indium content. The QW was formed by growing a 4nm layer of In${.81}$Ga${.19}$As, a 4nm layer of InAs, and a 10nm layer of In$_{.81}$Ga$_{.19}$As. A delta-doped Si layer was placed below the QW at a depth of 6nm, with a doping concentration of $n\sim 1 \times 10^{12}$ $cm^{-2}$. A schematic of the stack layers is shown in Fig.~\ref{fig:fig_1_cartoon}(a). These InAs quantum wells have been extensively studied in the context of mesoscopic superconductivity and topological superconducting states. Previous studies have mainly investigated the InAs/Al interface for developing tunable qubits and detecting topological superconductivity\cite{Larsen15, Casparis2018,2019Mayer_Mat, Ren2019, FornieriNature2019}. Recently the work has been extended to NbTiN~\cite{Mehdi2022} investigated the InAs/NbTiN interface in the context of semiconductor-superconductor heterostructures. In this new study, we aim to further explore this interface using QPCs to investigate the interplay between the integer quantum Hall effect (IQHE) and superconductivity with even greater precision.

\section{\textbf{III. Device fabrication and measurement setup}}

We fabricated a Hall bar using electron beam lithography and chemical wet etching. We cleaned the surface of the device using Argon plasma etching in the sputtering tool at 25 W power for 15 s followed by deposition of a 90 nm thick layer of NbTiN as the superconducting contacts. The interface between the InAs and NbTiN was 150~$\mu$m long, and a QPC was added with a separation of 150~nm between the QPC arms. A metallic gate for the QPC arms and pads was created by depositing 20 nm aluminum oxide (Al$_2$O$_3$) as the gate dielectric followed by 5~nm Cr and 20~nm Au e-beam deposition as the gate electrodes. Fig.\ref{fig:fig_1_cartoon}(b) shows the schematic of the device zoomed in around the QPC and interface area. The schematic demonstrates the edge modes reflection mechanism in IQH regime and when QPC is activated. Fig.\ref{fig:fig_1_cartoon}(b) shows an optical photo of the finished device labeling different region by their corresponding materials. 
We have fabricated two samples (A and B) to confirm our observations however we mainly focus on Sample A data in the main text and show all the data for sample B in the Supplementary Information (SI). 
The experiment was performed in a dilution fridge at a temperature of $T\sim 30$ mK with a maximum magnetic field of 12~T. The magnetotransport experiment was carried out using lock-in amplifiers and an AC four-point measurement technique with a frequency of $<20$ Hz and a $<1 \mu$A AC excitation current.

\section{\textbf{IV. Measurement results}}
\subsection{A. Magnetotransport data}
In order to assess the mobility of the quantum well (QW) in our fabricated device, we conducted magnetotransport experiments on the 2DEG. As shown in Fig.~\ref{fig:fig_1_cartoon}(c), we utilized contacts 1 and 4 as current source and drain, and measured R$_{xx}$(6-5) and R$_{xy}$(5-3) as a function of magnetic field for mobility and density analysis. Our measurements revealed that the mobility of the QW for sample A is approximately  $\mu \sim 13700$ $cm^{2}/V\cdot s$ at an electron density of $n = 1.54 \times 10^{12}$ $cm^{-2}$. This mobility value corresponds to an electron mean free path of approximately $l_e \sim 280 nm$. The data for $R_{xx}$ and $R_{xy}$ as a function of magnetic field is shown in Fig.~\ref{fig:fig_1_cartoon}(e), which indicates that the sample has a relatively high density. Several oscillations and plateaus in the longitudinal and Hall transport data are observed in the 12T window, respectively, corresponding to filling factors $\nu = 6,8$ and $10$. Fig.\ref{fig:fig_1_cartoon}(d) displays the downstream resistance $R_D$ (measured between contacts 5 and 4) as a function of QPC voltage and perpendicular magnetic field.

To understand the transport properties of mesoscopic systems, the Landauer-Büttiker formalism is widely used~\cite{buttiker1986}. When a superconductor is present, Andreev reflection (AR) processes can occur at the interface between the semiconductor and the superconductor~\cite{Hoppe2000}. To include the possibility of AR, we have extended the Landauer-Büttiker formalism describing edge state transport in a Hall bar geometry. 
%
The effect of the AR is captured by the average Andreev reflection $A\equiv (1/\nu)\sum_{i=1..\nu}A_i$, where $\nu$ is the total
number of edge modes, and $A_i$ is the probability of Andreev reflection, i.e., the probability of an electron-like chiral 
state approaching the QH-SC interface to be converted into a hole-like chiral state when leaving the QH-SC interface,~\cite{Mehdi2022}.
As $V_{QPC}$ is increased the distance of the QH edge modes from the QH-SC interface increases. 
As a consequence the strength of the superconducting correlations for the QH edge modes decreases, especially for the 
modes further removed from the edge, resulting in a decrease of $A$.

Following our previous work~\cite{Mehdi2022}, we begin by considering the six-terminal configuration shown in Fig.~\ref{fig:fig_1_cartoon}(b). The terminals 1, 2, 3, 5, and 6 are considered as ideal metallic leads, while contact 4 is a superconducting lead. 
Let $I_i$ and $V_i$ denote the currents and voltages, respectively, at the terminals $i = (1, 2,..., 6)$. Without loss of generality, we set $V_4 = 0$. The conservation of charge equation $\sum_i I_i = 0$ is employed to express $I_4$ in terms of the currents at the other leads. Using these considerations, the Landauer-Büttiker equations take the form
\begin{align}
\begin{split}
\begin{pmatrix} 
I_1 \\ I_2 \\ I_3 \\ I_5 \\ I_6 
\end{pmatrix} =& \frac{\nu}{R_H} 
\begin{pmatrix} 
1  & 0  & 0 & 0 & -1 \\ 
-1 & 1  & 0 & 0 & 0 \\ 
0  &  -1 & 1 & 0 & 0 \\ 
0  & 0  & 2A-1 & 1 & 0 \\ 
0  & 0  & 0 & -1 & 1 
\end{pmatrix}
\begin{pmatrix} 
V_1 \\ V_2 \\ V_3 \\ V_5 \\ V_6 
\end{pmatrix},
\end{split}
\label{eq:I_matrix}
\end{align}
where $\nu$ represents the total number of edge states, $R_H$, the Hall resistance ($h/e^2$), and A, the average probability of Andreev reflection per edge mode.

By assuming that no current flows into leads 2, 3, 5, and 6, we can simplify the equation by setting $I_2 = I_3 = I_5 = I_6 = 0$. We 
set the voltages at terminals 1, 2, and 3 equal to each other $V_1 = V_2 = V_3$, and the voltages at terminals 5 and 6 equal to each other $V_5=V_6$. Let $I=I_1= - I_4$. 
With these assumptions, we can  solve Eq.~\ref{eq:I_matrix} to obtain

\begin{align}
    & R_U = \frac{V_3 - V_4}{I} =  \frac{R_H}{\nu} \frac{1}{2A} \label{eq:R_U}\\
    & R_D = \frac{V_5 - V_4}{I} =  \frac{R_H}{\nu}\left(\frac{1}{2A}-1\right)\label{eq:R_D}
\end{align}
The set of equations outlined above provide a theoretical framework for studying edge state transport in the IQH/SC hybrid system. 

\begin{figure}[t]
\centering
\includegraphics[width=0.48\textwidth]{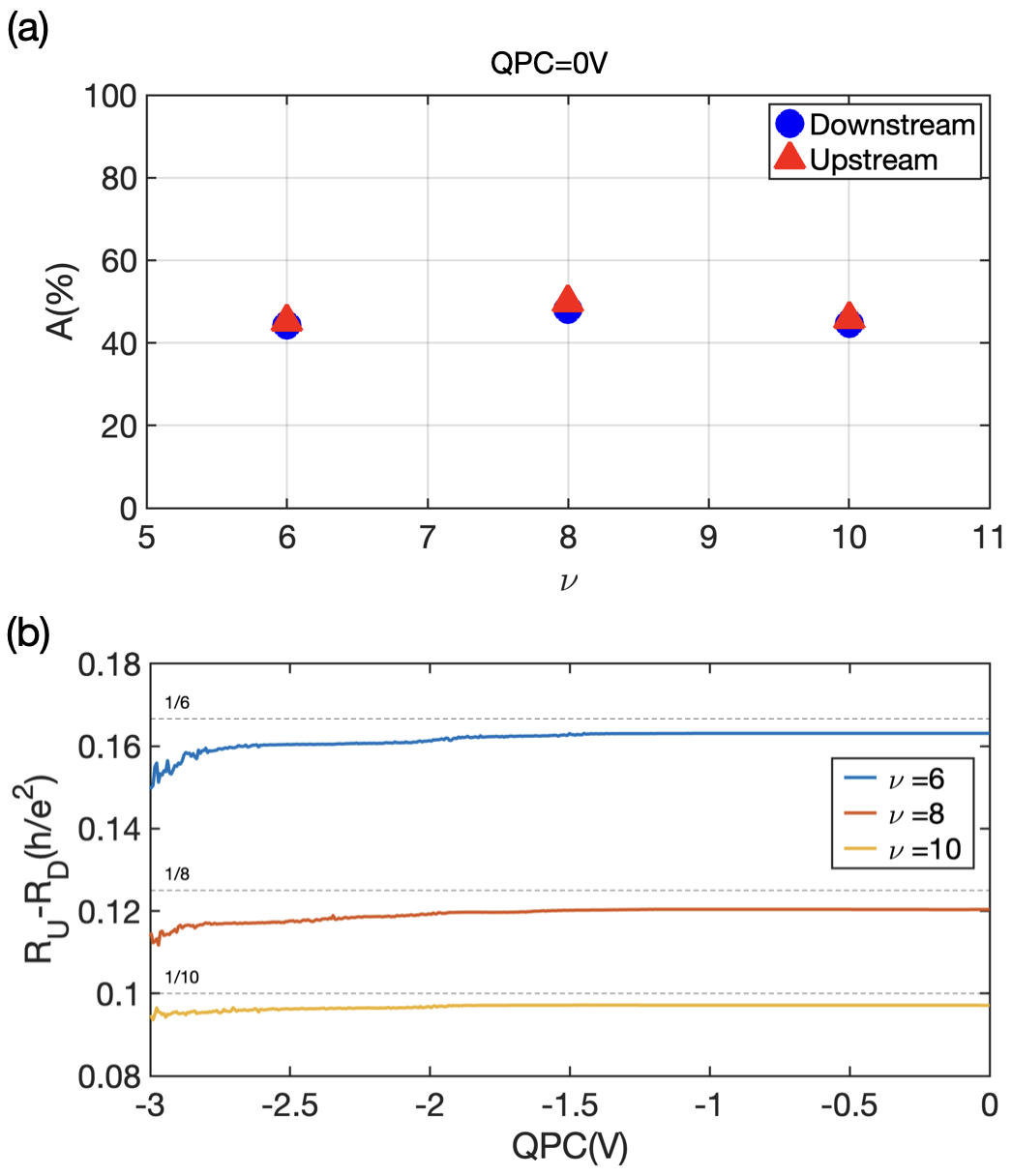}
\caption{(a) Extracted Andreev reflection data for three distinct filling factors: 6, 8, and 10. This data is derived from both the R$_D$ and R$_U$ datasets, and with a QPC voltage of 0V. (b) Dependence of the difference $R_U - R_D$ on QPC voltage.}
\label{fig:exp_QPC0}
\end{figure}

\begin{figure}[htp]
\centering
\includegraphics[width=0.49\textwidth]{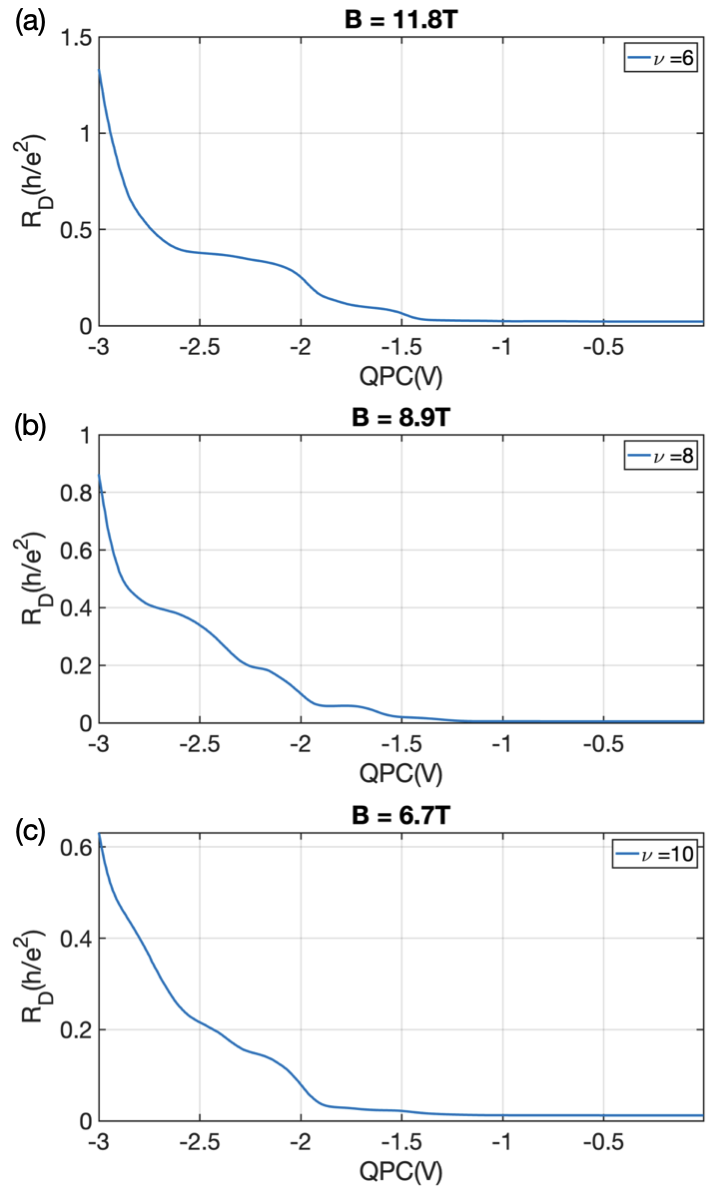}
\caption{Measured downstream resistance ($R_D$) as a function of QPC voltage for different values of $\nu$: (a) 6, (b) 8 and (c) 10. The magnetic field for each filling factor is labeled on top of each panel.}
\label{fig:exp_p_theo}
\end{figure}

Starting with the case where the QPC voltage is zero, we an extract $A$ using Eq.~(\ref{eq:R_U}-\ref{eq:R_D}) from $R_U$ and $R_D$. In Fig.~\ref{fig:exp_QPC0}(a) we show the extracted $A$ at different filling factors for sample A studied in this work. 
We see that for  $V_{QPC}=0$ the value of $A$ falls within the range of $40\%$--$50\%$
and that the  extracted value of $A$ is found to be consistent between upstream and downstream resistances.
Sample B shows a similar trend at higher filling factors filling factors ($\nu=12,~14,~16$) as shown in the SI. 

\begin{figure}[htb]
\centering
\includegraphics[width=0.48\textwidth]{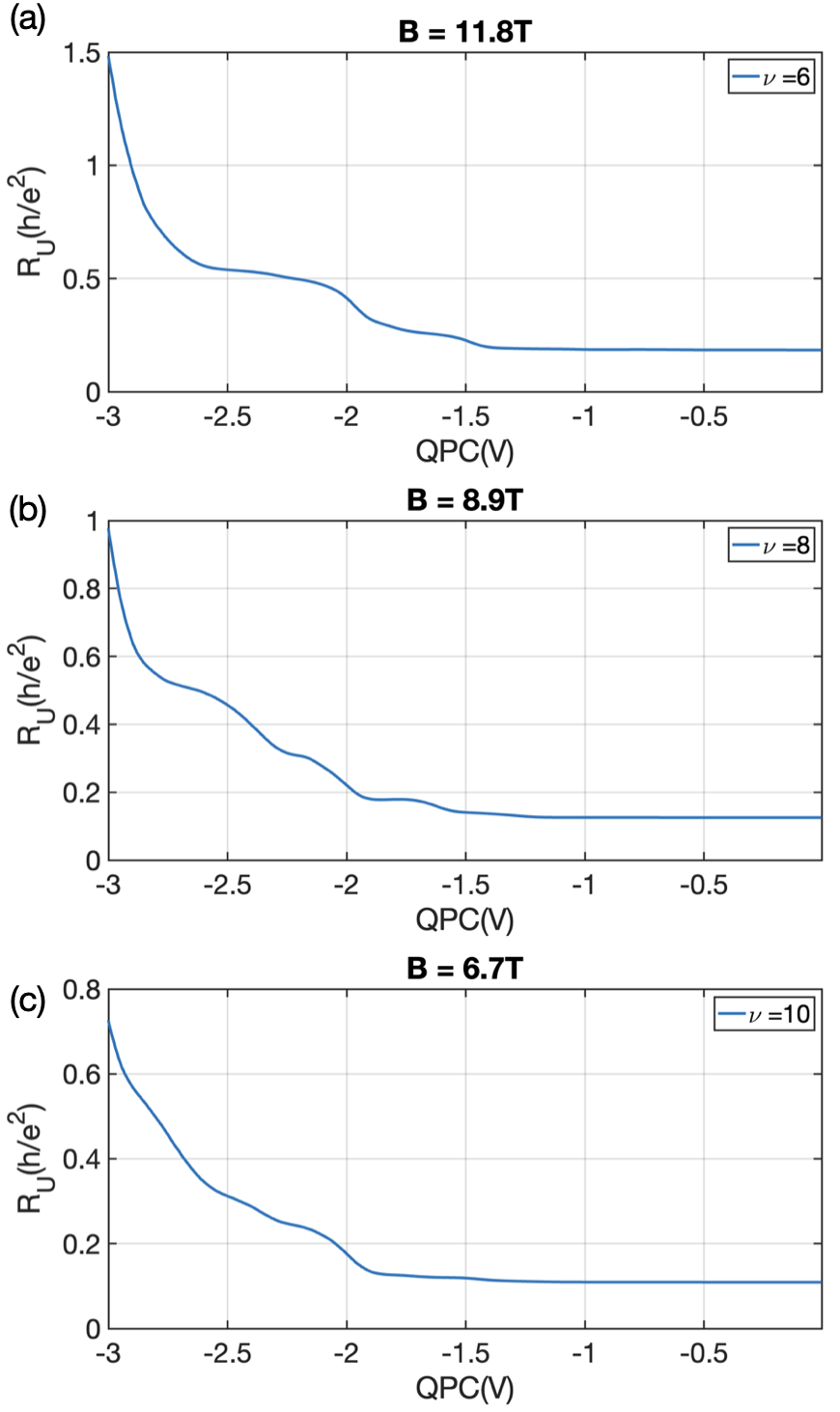}
\caption{Measured upstream resistance ($R_U$) as a function of QPC voltage for different values of $\nu$: (a) 6, (b) 8 and (c) 10. 
}
\label{fig:up_exp}
\end{figure}

Figure~\ref{fig:exp_p_theo} shows the downstream resistance as a function of QPC voltage for different fillings ($\nu = 6,~8,~10$). In each case, the downstream resistance remains constant for voltages $\gtrsim -1.5$~V before increasing. 
In Fig.~\ref{fig:up_exp}, we present the upstream resistance at the same magnetic fields and QPC range as Fig.~\ref{fig:exp_p_theo}. There we also observe an increase in the resistance with decreasing QPC voltage for $V_{QPC}< -1.5$~V.
We can check the consistency of Eqs.~(2-3) with the experimental measurements by considering the difference 
$(R_U - R_D)$. According to Eq.(\ref{eq:R_U}-\ref{eq:R_D}), $(R_U - R_D) = h/(\nu e^2)$ regardless of the transmission through the QPC.
From Fig.~\ref{fig:exp_QPC0}(b) we see the difference in resistances remains fairly constant with QPC voltage $> -3$~V and corresponds to the expected filling. This suggests that Eqs.~(\ref{eq:R_U}-\ref{eq:R_D}) consistently model the experimental data between the two edge state transport regimes
for $V_{QPC}>-3$~V. 
When the QPC voltage falls below $-3$~V, the R$_U$ and R$_D$ quickly diverges and the Landuaer-B\"{u}ttiker model breaks down.

\begin{figure}[t]
\centering
\includegraphics[width=0.48\textwidth]{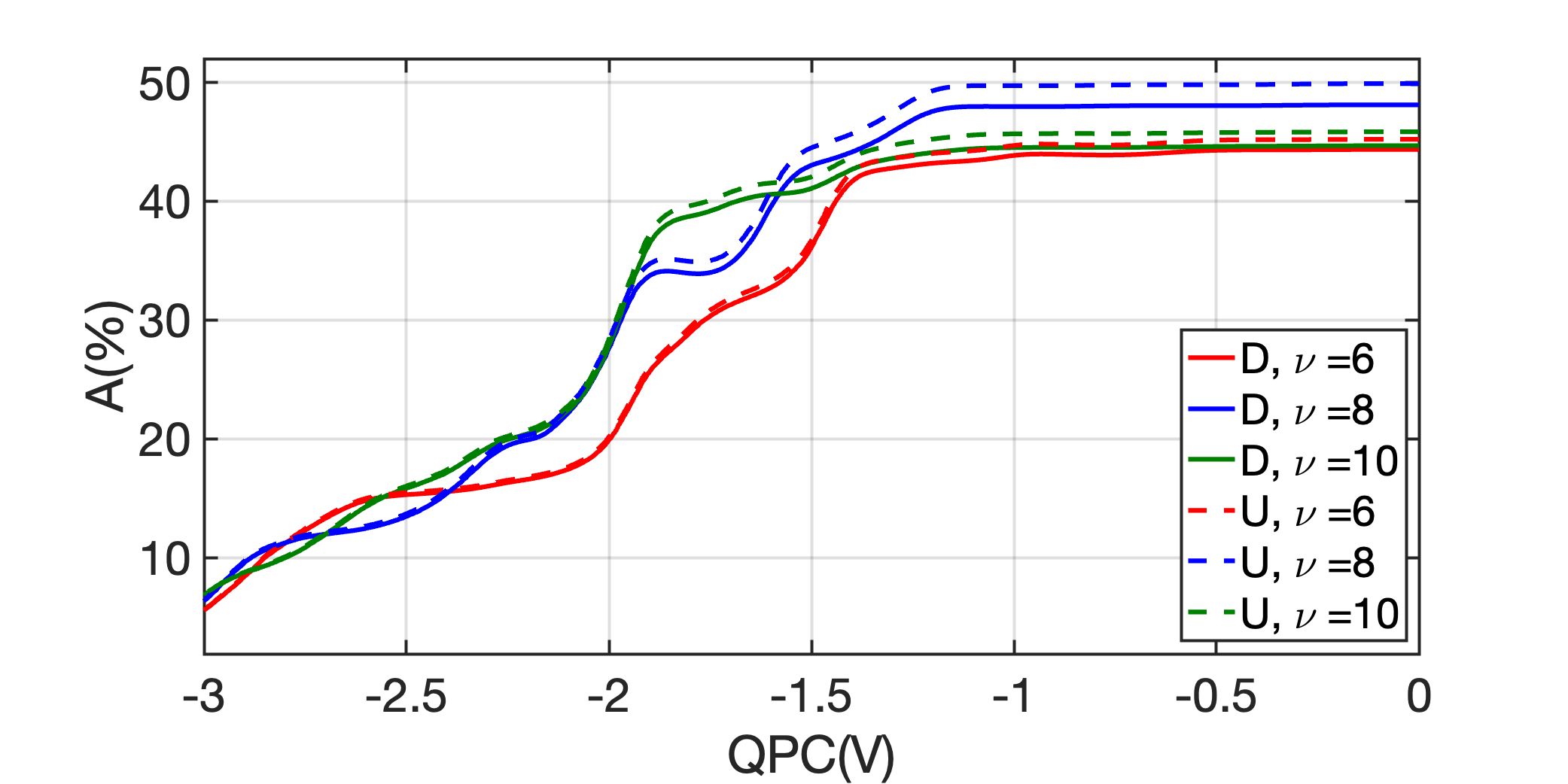}
\caption{The extracted $A$ parameter from both upstream (U) and downstream (D) resistance for different filling factors $\nu$ as a function of QPC voltage.}
\label{fig:fig5_A}
\end{figure}
From the experimental measurement of $R_D$ and $R_U$ shown in Figs.~\ref{fig:exp_p_theo},~\ref{fig:up_exp},
using Eqs.(2)~(3) we can obtain the evolution of $A$ with $V_{QPC}$ for different filling factors $\nu$.
Figure~\ref{fig:fig5_A} shows the obtained average Andreev conversion probability.
We see that, regardless of the bulk filling factor, $A$ is suppressed by the QPC. 
We also observe that for $V_{QPC}<-1$~V, $A$ first decreases continuously and then, for some ranges
of $V_{QPC}$, exhibits plateaus, as can be seen most clearly for the case $\nu=8$ when $-2<V_{QPC}<-1.7$~V. 
The continuous decrease of $A$ with $V_{QPC}$ can be interpreted as due to the progressive
reduction of the superconducting pairing correlations induced by the superconductor into the QH edge modes
as the QPC pushes away the edge modes from the QH-SC interface. As $V_{QPC}$ reaches threshold values
some of the edge modes further removed from the edge are almost completely pinched off from the QH-SC interface
resulting in a new, reduced, value of $A$ that approximately is not affected by a further decrease of $V_{QPC}$
until $V_{QPC}$ is large enough, in absolute value, to significantly affect the pairing correlations of one more edge mode. Let the number of edge states remaining be $\nu_s$.
Then for the ranges of $V_{QPC}$ for which $A$ is approximately constant 
we can assume that $\nu_s<\nu$ modes have a value of $A\neq 0$, while $\nu-\nu_s$ modes have $A\approx 0$.
One of the effects of the QPC is to also reduce the effective length, $L_{sc}$, of the QH-SC interface.
When $V_{QPC}=0$ we can assume $L_{sc}=L=150 \mu$m. For the largest absolute value of $V_{QPC}$ 
$L_{sc}$ can be taken to be equal to the distance between the two gates forming the QPC: $L_{sc}=150~$nm.
In both cases $L_{sc}$ is much greater than the superconducting coherence length of NbTiN ($\sim 10$ nm) so that crossed Andreev reflection processes can be neglected~\cite{Lee2017, Gul2022}.
However, one would expect that the variation of $L_{sc}$ induced by $V_{QPC}$ would induce oscillations
in $A$. The lack of oscillations of $A$ with $V_{QPC}$ suggests
that effects due to disorder and vortices in NbTiN might play an important role
resulting on an effective averaging of $A$ over $L_{sc}$, as proposed in Ref.~\cite{Mehdi2022},
giving rise to a value of $A$ averaged along the length $L_{sc}$ that is independent of $L_{sc}$
as long as $L_{sc}$ is much large than the superconducting coherence length $\xi$ of the superconductor.

\section{\textbf{V. Conclusion}}
Our work demonstrates the successful fabrication of a hybrid device at the InAs/NbTiN interface, incorporating a quantum point contact (QPC). 
We have shown that by tuning the QPC voltage the effective, average, Andreev conversion probability $A$
for QH edge modes can be tuned. We find that there are threshold values of the QPC voltage
for which some of the QH edge modes appear to be completely pinched off by the QPC from
the QH-SC interface resulting in plateaus in the scaling of $A$ with $V_{QPC}$. 
The results also show that the variation induced by the QPC of the effective length $L_{sc}$ of the QH-SC interface
does not result in oscillations of $A$. This is consistent
with the findings of Ref.~\cite{Mehdi2022} and suggests that
effects due to disorder and vortices in NbTiN must play an important role
in determining the properties of chiral Andreev states in InAs/NbTiN QH-SC heterojunctions 
resulting on an effective averaging of $A$ over $L_{sc}$ that is independent of $L_{sc}$
as long as $L_{sc}$ is much larger than the superconductor's coherence length. These findings advance the understanding of QH-SC interfaces and should motivate future works to further elucidate the 
details of the interplay of QH and superconducting states.
 
\section{\textbf{Acknowledgements}}
This work was supported by US DOE BES Grant No DE-SC0022245 and U.S. Army Research Office Agreement No. W911NF1810067.
The work at Sandia is supported by a LDRD project. 

Sandia National Laboratories is a multi-mission laboratory managed and operated by National Technology \& Engineering Solutions of Sandia, LLC (NTESS), a wholly owned subsidiary of Honeywell International Inc., for the U.S. Department of Energy’s National Nuclear Security Administration (DOE/NNSA) under contract DE-NA0003525. This written work is authored by an employee of NTESS. The employee, not NTESS, owns the right, title and interest in and to the written work and is responsible for its contents. Any subjective views or opinions that might be expressed in the written work do not necessarily represent the views of the U.S. Government. The publisher acknowledges that the U.S. Government retains a non-exclusive, paid-up, irrevocable, world-wide license to publish or reproduce the published form of this written work or allow others to do so, for U.S. Government purposes. The DOE will provide public access to results of federally sponsored research in accordance with the DOE Public Access Plan.
ER thanks KITP, supported in part by the National Science Foundation under Grants No. NSF PHY-1748958 and PHY-2309135,
where part of this work was performed.
\begin{center}
{\bf References}
\end{center}

\bibliography{References_Shabani_Growth}
\pagebreak

\end{document}